\documentclass[aps,graphicx,twocolumn,epsfig]{revtex4}
\usepackage{amsmath}
\usepackage{graphicx}

\usepackage{graphicx}
\usepackage{dcolumn}
\usepackage{bm}

\newcommand{\ud}{\mathrm{d}}

\newcommand{\be}{\begin{equation}}
\newcommand{\ee}{\end{equation}}

\begin{document}


\title{Phase Sensitivity of a Mach-Zehnder Interferometer}

\author{Luca Pezz\'e and Augusto Smerzi}
\affiliation{ Theoretical Division, Los Alamos National Laboratory,
Los Alamos, New Mexico 87545, USA\\
and\\
Istituto Nazionale per la Fisica della Materia BEC-CRS \& Dipartimento di Fisica, Universit\`a di Trento, I-38050 Povo, Italy}

\date{\today}

\begin{abstract}
The best performance of a Mach-Zehnder interferometer is achieved with the
input state 
$|N_T/2 + 1\rangle |N_T/2-1 \rangle+|N_T/2 - 1\rangle|N_T/2+1\rangle$, being
$N_T$ the total number of atoms/photons.
This gives:  
i) a phase-shift error confidence $C_{68\%}=2.67/N_T$ with ii) a $single$ 
interferometric measurement.  
Different input quantum states can
achieve the Heisenberg scaling $\sim 1/ N_T$ but with higher prefactors 
and at the price of a statistical analysis 
of {\it two or more} independent measurements.  
\end{abstract}

\maketitle

\begin{figure}[!h]
\begin{center}
\includegraphics[scale=0.5]{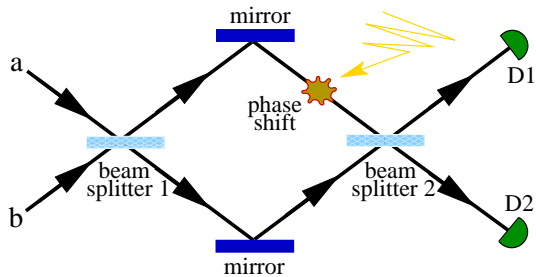}
\end{center}
\caption{\small{Schematic representation of the Mach-Zehnder interferometer. Atoms/photons enter the $a$
and $b$ input ports, mix and recombine in the beam splitters and are finally detected in D1 and D2. The 
phase shift is inferred from the number of atoms/photons measured in each output port. }}
\label{mz} 
\end{figure}

{\it Introduction.} 
The central goal of interferometry is
to estimate phase shifts with the highest sensitivity
given a finite resource of atoms/photons.
There are several possible interferometric configurations, which are generally reducible 
to the ``drosophila'' Mach-Zehnder (MZ), see Fig.(\ref{mz}). 
The MZ transforms a two-mode input state as $|\psi \rangle_{out} = e^{- i \hat{J}_y \theta} |\psi \rangle_{inp}$ \cite{Yurke_1986}, 
where the generator of the phase translation $\theta$ is 
the $y$ angular momentum component of the three-dimensional rotation group 
$\hat{J}_x=(\hat{a}^{\dag}\hat{b}+\hat{a}\hat{b}^{\dag})/2$, 
$\hat{J}_y=(\hat{a}^{\dag}\hat{b}-\hat{a}\hat{b}^{\dag})/2i$, 
$\hat{J}_z=(\hat{a}^{\dag}\hat{a}-\hat{b}^{\dag}\hat{b})/2$ 
($\hat{a},\hat{b}$ are bosonic annihilation operators for the two input quantum modes).
How precisely can the unknown phase shift $\theta$ be estimated from the output $|\psi \rangle_{out}$ ?
It is well known that,
if one of the two input arms of the interferometer is fed with vacuum, the phase measurement uncertainty  
is bounded by the standard quantum limit (or ``shot noise'') 
$\Delta \theta \ge 1/\sqrt{N_T}$, where $N_T $ is the total/average number of atoms
in the input state.
The first breakthrough came in the early '80, when Caves showed that 
it is possible to reach 
sub shot-noise sensitivities after squeezing the vacuum of the unused port of the interferometer
\cite{Caves_1981}. Sub shot-noise was experimentally 
demonstrated few years later \cite{Xiao_1987}. 
This inaugurated 
a large body of literature proposing new states and optimal performances \cite{Yurke_1986, Yurke_1986_b, Holland_1993, Hillery_1993,
Bollinger_1996,ariano2000} to  push the sensitivity up to the Heisenberg limit $\sim 1/N_T$ \cite{Luis_2000}.
Paradoxically, however, fewer efforts have been devoted to the correct inference of phase measurement uncertainties.
This fact went (and it is still going) almost unnoticed, despite of the seminal 
works of Helstrom and Holevo \cite{Helstrom} and Lane, Braunstein and Caves \cite{Lane_1992}. 
Apart from a few noticeable exceptions, in the current literature
the phase measurement uncertainty is still retrieved from a simplified error propagation theory,
$\Delta \theta=(\Delta \hat{A})/\big|\frac{\partial \langle \hat{A} \rangle}{\partial \theta} \big|$, where
$\langle \hat{A} \rangle$ is the average of a generic phase dependent observable $\hat{A}(\theta)$
and $\Delta \hat{A} = \langle \hat{A}^2 \rangle - \langle \hat{A} \rangle^2 $ is the mean square fluctuation.
Unfortunately, this simple approach can give correct estimates only when the probability
distributions are Gaussian \cite{nota5}, which is seldom the case in quantum interferometry.
Moreover, optimized phase estimations often require
the distribution of the available finite resource of $N_T$ particles among 
$p$ {\it independent} interferometric experiments, each with $N=N_T/p$ particles.
This optimization cannot be analyzed with the linear error propagation formula, which
only considers the dependence on the number of particles in each experiment $N$ and not on 
the total number of particles $N_T$.
As a matter of fact, even if the Mach-Zehnder has been extensively studied,
there is not published literature about the exact dependence of phase uncertainties on the
total (or mean) number of particles of any input state. 

Matter waves Mach-Zehnder interferometers have been realized 
with cold atoms \cite{Gustavson_2000} and electrons \cite{Yang_2003}.
Evidences of non-classical light properties after a beam splitter have been experimentally showed in \cite{Feng_2004}.
Entangled ``NOON'' states have been created with few photons \cite{Mitchell_2004}
and ions \cite{Leibfried_2004}
with applications in quantum lithography \cite{Boto_2000}.
Current efforts are devoted to the creation of the MZ with dilute Bose-Einstein condensates (BEC). 
BEC offers almost unique possibilities
to create the large intensity, highly squeezed waves needed for Heisenberg interferometry.
Recently, double-slit \cite{Shin_2004} as well as Michelson interferometers \cite{Wang_2005} 
have been experimentally demonstrated. By trapping
BEC in two wells and in periodic potentials, the beam splitters and mirrors  
are replaced by the dynamical tailoring of interwell barriers and magnetic wells.
Fock and number squeezed condensates have been realized in \cite{Greiner_2002}. 

In the following, we study a MZ interferometer fed by
\be \label{general}
| \psi \rangle_{inp} =\frac{1}{\sqrt{2}}\big( |j+m \rangle_a |j-m \rangle_b + |j-m \rangle_a |j+m
\rangle_b \big),
\ee
where $N=2 j$ and $2 m$ are the total and the relative number of particles at the $a$ and
$b$ input ports, respectively, cfr. Fig. (\ref{mz}). 
We rigorously calculate error confidences within a Bayes framework \cite{Helstrom}. 
We first consider the twin-Fock ($m=0$) introduced by Holland and Burnett \cite{Holland_1993}.
We then study a ``NOON'' state ($m=j$), showing that it cannot overcome shot-noise sensitivity.
We will finally demonstrate that the best Mach-Zehnder performance is obtained when
$m=1$. This state 
can be created by
constructing a twin-Fock state followed by the measurement of one particle \cite{Castin_1997}.

{\it Bayes analysis}.
Quantum Mechanics provides the conditional probability $P(\mu|j, \theta)$
to measure, at the output ports, the relative number of atoms $\mu = (N_1 -N_2)/2$, given $2j=N_1+N_2$ particles 
and the unknown phase shift $\theta$. 
With the general state Eq.(\ref{general}) we have
\begin{equation}
P(\mu|j,\theta)=|\langle j, \mu|\psi \rangle_{out}|^2= \frac{1}{2} | d^j_{\mu,m}(\theta)+d^j_{\mu,-m}(\theta)|^2, 
\label{pmutheta}
\end{equation}
where $| j, \mu \rangle \equiv |j + \mu \rangle_{D1} |j - \mu \rangle_{D2}$, being D1 and D2 the detectors at the output
ports (see Fig. (\ref{mz})), and $d^j_{\mu, \nu}(\theta)$ being the angular momentum matrices \cite{nota1}. 
The goal is to estimate $\theta$ after detecting a certain value $\mu$.
According to the Bayes theorem,
$P(\theta|j, \mu)P(\mu)= P(\mu|j, \theta) P(\theta)$, where 
$P(\theta)$ and $P(\mu)$ take into account any further {\it a priori} information about the real value of the phase shift
and the output measurement $\mu$. We assume complete ignorance, so $P(\theta)=\mathrm{H}\big( \pi/2-|\theta| \big)$, 
being $\mathrm{H}(x)$ the Heaviside step function, while $P(\mu)$ is fixed by the normalization.
We notice that, with the state Eq.(\ref{general}) as input of the MZ, it is possible to estimate only the 
absolute value of phase shifts 
in the interval $[0, \pi/2]$.
In most cases, the best interferometric performance  requires the statistical analysis of
several independent measurements.
Once obtained the results $\mu_1 ... \mu_p$, the phase probability distribution reads
$P_p(\phi|j, \mu_1 ... \mu_p) \propto \prod_{k=1}^{p} P(\phi|j,\mu_k)$.
Averaging over all the possible $p$-uple $ \mu_1 ... \mu_p$ we find the phase 
distribution for fixed $\theta$ and $j$:
\be \label{phasedist}
P_p(\phi|j,\theta)=\sum_{ \{\mu_1 ... \mu_p\}} P_p(\phi|j, \mu_1 ... \mu_p) P_p(\mu_1 ... \mu_p|j, \theta),
\ee
where $P_p(\mu_1 ... \mu_p|j, \theta)= \prod_{k=1}^{p} P(\mu_k| j, \theta)$ is provided by quantum mechanics.
As phase estimator, we choose the value of the phase $\bar{\phi}_p(j, \theta)$ corresponding to the maximum
of the probability distribution Eq.(\ref{phasedist}). 
The phase uncertainty is estimated as confidence, namely the $\gamma$-probability 
that the real value of the phase shift is within a given interval $c_{\gamma}$ around
in  $\bar{\phi}$:
$\gamma=\int_{\bar{\phi}-c_{\gamma}}^{\bar{\phi}+c_{\gamma}} \ud \phi \, P_p(\phi|j, \theta )$.

In the following we consider the case of a null phase shift $\theta=0$.
The probability distribution Eq.(\ref{pmutheta})
reduces to $P(\mu|j,0)=|\delta_{\mu,m}+\delta_{\mu, -m}|^2/2$, where $\delta_{\alpha,\beta}$ is 
the Kronecker delta. 
Therefore, we have two possible results for the measurement of the relative 
number of particles: $\mu=+m$, and $\mu=-m$. Because of  
the symmetry properties of the MZ interferometer and of the state Eq.(\ref{general}), we have that
$P(\phi|j,m)=P(\phi|j,-m)$, and the sum Eq.(\ref{phasedist}) reduces to 
$P_p(\phi|j,0)\approx | d^j_{m,m}(\phi)+d^j_{m,-m}(\phi)|^{2p}$.
When $\theta\neq 0$ the probability of obtaining values 
of $\mu \neq \pm m$ increases. In this case we might need several measurements 
in order to explore all the relevant probability
distribution, so the sensitivity 
of the interferometer for a given number of particles is, in general, reduced \cite{Sanders_1995}.

\begin{figure}[t]
\begin{center}
\includegraphics[scale=0.62]{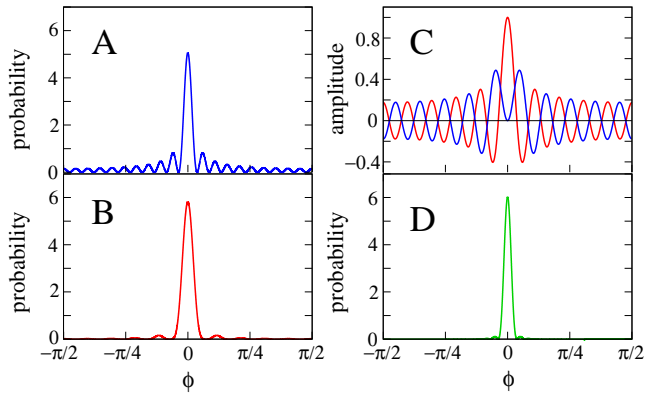}
\end{center}
\caption{\small{ 
A,B: Probability distribution of the $m=0$ twin-Fock state with A)
$N=40$ particles and a $p=1$ measurement, (A); 
$N=20$ particles and $p=2$ measurements, (B). 
By combining the independent measurements it is possible to strongly reduce the weight of the tail of the distribution
with respect to the central peak.
C,D: Amplitude $d^j_{1,1}(\phi)$ (red line), and $d^j_{1,-1}(\phi)$ (blue line), 
for $N=40$ particles ($j=20$) \cite{nota1}, (C). The tails of the amplitudes oscillate 
out of phase and interfere destructively, so to enhance the central peak of the 
probability distribution $P(\phi|j, 0)\sim|d^j_{1,1}(\phi)+d^j_{1,-1}(\phi)|^2$, (D). }}\label{distphi} 
\end{figure}

{\it Mach-Zehnder interferometer with twin-Fock States}.
We first consider the input state $|\psi \rangle_{inp} =|j\rangle_a |j \rangle_b$ ($m=0$ in Eq.(\ref{general})).

The probability of measuring a relative particle number $\mu$ at the output port, given a phase shift $\theta$,
is \cite{Kim_1998}:
$P(\mu|j, \theta) = \frac{(j-\mu)!}{(j+\mu)!}\big[
P_j^{\mu}(\cos \theta) \big]^2$,
where $P_j^{\mu}(x)$ are the Associated Legendre Polynomials. 
When $\theta=0$ the probability $P(\mu|j, \theta)$ reduces to a single peak centered in $\mu=0$. 
Therefore, repeating the experiment $p$ times we have
$ P_p(\phi|j, 0) \approx [P_j^0(\cos \phi)]^{2p}$. 
With a single measurement ($p=1$), the $68\%$-confidence scales as $\sim 1/N_T^{1/3}$ while, as discussed in 
\cite{Hradil_2005}, the mean square phase fluctuation scales as $\sigma \sim 1/\ln N_T$. In both cases 
the phase uncertainty is worse than shot noise \cite{nota3}.
This problem is generally overlooked when the sensitivity is calculated from the simplified error propagation formula, 
or looking at the width of the central 
peak of the Legendre polynomials, which indeed shrink as $1/N_T$. 
The large tails of the distribution contain most of the probability and cannot be ignored, see Fig.(\ref{distphi},A).
This clarifies the need of considering several independent measurements, which increases the relative weight
of the central peak with respect to the tails, see Fig. (\ref{distphi},B).
As discussed in \cite{Hradil_2005}, the mean square phase fluctuation is minimized 
combining $p=4$ measurements.

\begin{figure}[t]
\begin{center}
\includegraphics[scale=0.72]{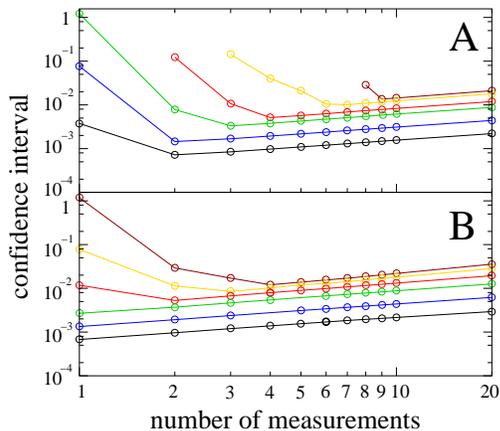}
\end{center}
\caption{\small{Confidence interval as a function of the number of independent measurements
$p$ for A) the twin-Fock $m=0$ and 
B) the twin $m=1$ states. Here $N_T=2000$ particles. In both plots, from the bottom  line to the top one, 
the black line is the confidence at $38.29 \%$, the blue at $68.27 \%$; 
the green at $95.45\%$, 
the red at $99.73 \%$, the yellow at $99.994 \%$,
and the brown at $99.99994 \%$. These correspond to $\sigma/2, \sigma, 2 \sigma, 3 \sigma, 4 \sigma$ and $5 \sigma$
when the distribution is Gaussian.}} 
\label{confidenza} 
\end{figure}

\begin{figure}[t]
\begin{center}
\includegraphics[scale=0.72]{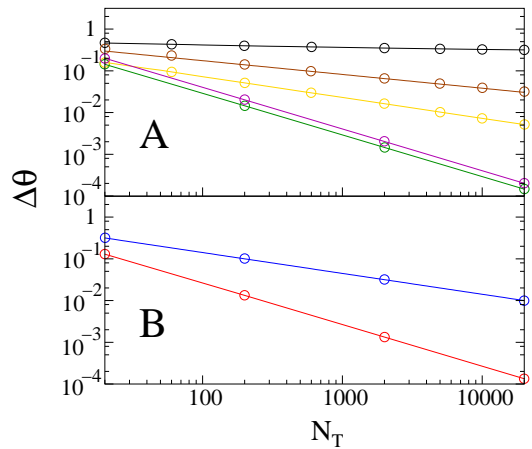}
\end{center}
\caption{\small{Phase uncertainty $\Delta \theta$ as a function of $N_T$. A) Twin-Fock state: 
(from the top line to the bottom one)  black line, $\sigma$ for $p=1$ ($\sigma \approx 1/\ln N_T$);
brown line, $68 \%$ confidence for $p=1$ ($\Delta \theta= 0.8/N_T^{1/3}$);
yellow line, $\sigma$ for $p=2$ ($\sigma \approx 1/N_T^{1/2}$);
violet line, $\sigma$ for $p=4$ ($\sigma = 4/N_T$);
green line: $68 \%$ confidence for $p=2$ ($\Delta \theta= 2.91/N_T$).
B) Phase sensitivity of the NOON state (blue line), $\sigma = \sqrt{2/N_T}$, independent of $p$,
and of the twin $m=1$ state (red line), $\Delta \theta=2.67/N_T$ at $p=1$. Points are numerical data.}} 
\label{scaling} 
\end{figure}

In Fig. (\ref{confidenza},A) we study different values of confidence for the twin-Fock state 
distribution as a function of the number $p$ of independent measurements and a fixed total number of
atoms $N_T$.
We see that the bigger the confidence, the larger is the number of 
experiments, $p$, necessary to reach the minimum. 
It is important to notice that the number of measurements needed to reach the minimum for a given 
confidence does not depend on the total number of atoms $N_T$. 
The main results of our analysis can be summarized as follows: 
i) with a single measurement the MZ interferometer gives a sensitivity worse than the shot noise, 
ii) the highest sensitivity is reached with {\it two} independent
measurements, for the $68\%$ confidence, and iii) with {\it three} measurements
for the $95 \%$ confidence:
\be
\Delta \theta_{68\%}=\frac{2.915}{N_{T}},~~~~p=2; \qquad \Delta \theta_{95 \%}=\frac{6.654}{N_{T}},~~~~p=3.
\ee
The number of measurements needed to reach higher confidences is shown in Fig.(\ref{confidenza},A). 
In Fig.(\ref{scaling},A) we summarize the scaling of the phase uncertainties reached with the 
twin-Fock state, as a function of $N_T$.

How does the sensitivity scale with the number of measurements, $p$, given a fixed number of atoms, $N$, 
for each experiment? The Cramer-Rao theorem does not allow a better scaling than $1/\sqrt{p}$, 
and we have verified numerically that the limit is reached when $p>4$, 
obtaining the mean square fluctuation $\Delta \theta \approx 2 \sqrt{p}/N_{T}$.
The $1/p$ scaling claimed in \cite{Kim_1998} was probably an artifact of low statistics.

{\it Mach-Zehnder interferometer with NOON state.}
We now consider the state (\ref{general}) with $m=j$.
If $\theta=0$ the only possible outcomes are 
$\mu=\pm j$. According to the Bayes theorem, we obtain the probability distribution
$P_p(\phi|j, 0) \approx [\cos^{2j}(\phi/2) + \sin^{2j}(\phi/2)]^{2p}$.
For large $j$ we have that 
$\cos^{2j}(\phi/2)\sim \exp[-(j/4)\phi^2]$, while 
$\sin^{2j}(\phi/2)\sim \exp[-(j/4)(\phi-\pi)^2]$. 
Therefore, the phase probability distribution is a Gaussian with width scaling as
$\sigma=\sqrt{2}/\sqrt{N_T}$ (see Fig.(\ref{scaling},B)).
The NOON state with a Mach-Zehnder interferometer, assuming a complete ``a priori'' ignorance of the
real phase shift, achieves a shot noise sensitivity.
A previous analysis \cite{Pezze_2005}, directly addressing current experiments \cite{Mitchell_2004, Leibfried_2004}, 
showed that NOON states 
and balanced homodyne detection can provide phase sensitivity scaling as $\sim 1/N_T^{3/4}$.

{\it Mach-Zehnder interferometer with a twin $m=1$ state.} 
We have previously shown how the large tails of the twin-Fock probability 
prevent the sensitivity to scale as 
$\sim 1/N_T$. To kill these tails, and enhance the central peak of the distribution, we have to 
statistically analyze several independent measurements. 
Here we follow a different strategy: is it possible to cancel out the tails by 
destructive interference?
The answer is positive. 
The tails of the probability distribution $P(\phi|j, 0)\sim|d^j_{m,m}(\phi)+d^j_{m,-m}(\phi)|^2$ 
obtained with the input state
(\ref{general}) cancel out when $m$ is odd. This special feature, not present in
the twin-Fock state, is a consequence of the interference 
between the two matrices in $P(\phi|j,0)$, and
it can be easily recognized by expanding the Jacobi Polynomials \cite{nota4}.
Since the cancellation has a precision of the order of $O[(j-m)^{-3}]$ we can conclude that,
among this class of states, the case $m=1$ offers the highest performance.
States with larger odd values of $m$ do not benefit of an equally efficient destructive 
interference. With semi-integer values of $m$, as for the Yurke state $(m=1/2)$ \cite{Yurke_1986_b}, 
there is no cancellation at all, so we expect that a $1/N_T$ scaling can be reached only with $p>1$ as
in the twin-Fock case. 
In Fig.(\ref{distphi},C) we plot the amplitude $d^j_{1,1}(\phi)$ (red line), and 
$d^j_{1,-1}(\phi)$ (blue line), for $N=40$ particles ($j=20$). Outside the central
region the two functions oscillate out of phase and, when summed   
to calculate $P(\phi|j, 0)$, interfere destructively. The phase probability distribution is shown in 
Fig.(\ref{distphi},D).
The comparison with the Fock state distribution (Fig. (\ref{distphi},A)), for the same total number of particles, highlights the
strong reduction of the tails. As a consequence, for a given confidence, we reach 
the minimum with a smaller number of 
measurements with respect to the Fock state case, see Fig. (\ref{confidenza},B). 
The sensitivity of the twin-$m$ states is:
\be
\Delta \theta_{68\%} =\frac{2.67}{N_{T}},~~~~p=1; \qquad\Delta \theta_{95\%} = \frac{5.376}{N_{T}},~~~~p=1.
\ee
The scaling of $\Delta \theta_{68\%}$ as a function of $N_T$ is shown in Fig.(\ref{scaling},B).
The main advantage is, apart from the smaller prefactors,
that the $1/N_{T}$ scaling is reached with a single ($p=1$) measurement for
both the $68 \%$ and the $95 \%$ confidence, as compared with the $p=2$ and $p=3$ measurements requested
with Fock states, respectively. For higher confidences, the required number of measurements is shown in
Fig.(\ref{confidenza},B) \cite{nota10,notaf}.

{\it Conclusions}.
There are not known protocols for the direct measurement of phase shifts. 
Phases can only be inferred after the measurement of a different (phase dependent) 
observable. What is the highest achievable sensitivity? 
Within a rigorous Bayesian analysis we find that
$|\psi\rangle_{inp} = 1/\sqrt{2}
(|N_T/2 + 1\rangle |N_T/2-1 \rangle+|N_T/2 - 1\rangle|N_T/2+1\rangle$), 
gives the \emph{highest sensitivity} $C_{68\%}=2.67/N_T$, with a \emph{single}
interferometric measurement.
Different input quantum states can
achieve the Heisenberg scaling $\sim 1/ N_T$ but with higher prefactors
and at the price of a statistical analysis
of {\it two or more} independent Mach-Zehnder measurements.

{\it Acknowledgment}. We thank O. Pfister and B.C. Sanders for useful 
discussions.


\begin{thebibliography}{99}
\bibitem{nota5}
With non-Gaussian probability 
distributions, $\Delta \langle\hat{A}\rangle$ cannot be approximated by $\Delta \hat{A}$ in a error propagation
approach. Indeed, according to the central limit theorem, the variance of the mean, $(\Delta \langle \hat{A}\rangle)$, 
becomes equal to the average of the variance, $(\Delta \hat{A})/\sqrt{p}$ only asymptotically with
the number of independent measurements $p$.
\bibitem{nota3} We emphasize that when $\theta=0$, the Bayesian analysis of the measurement 
of the parity operator \cite{Campos_2003} or of the $\hat{J}_z^2$ operator \cite{Bouyer_1997}
give the same results as the measurement of the relative number of particles.
\bibitem{nota10}
The possibility to estimate a phase with a single measurement   
has some important advantages. Indeed, 
when the external perturbation has a
short life-time, the need to perform 
several independent measurements might
require the building of an equal number of interferometers.
\bibitem{notaf} We have extended our analysis to consider classes of states different from 
Eq.(\ref{general}). We have always found 
that the twin $m=1$ state provides the best performance, and, on the ground of some general arguments, 
we believe that this remains true with any possible two-mode input quantum state.
\end{thebibliography}
\end{document}